\documentclass[aps,nofootinbib,preprint,superscriptaddress]{revtex4}%
\usepackage{hyperref}
\usepackage{amsmath}
\usepackage{amsfonts}
\usepackage{amssymb}
\usepackage{graphicx}
\usepackage{color}%
\providecommand{\U}[1]{\protect\rule{.1in}{.1in}}
\begin{document}
\title{Spin polarization independence of hard polarized fermion string scattering amplitudes}
\author{Sheng-Hong Lai}
\email{xgcj944137@gmail.com}
\affiliation{Department of Electrophysics, National Chiao-Tung University, Hsinchu, Taiwan, R.O.C.}
\author{Jen-Chi Lee}
\email{jcclee@cc.nctu.edu.tw}
\affiliation{Department of Electrophysics, National Chiao-Tung University, Hsinchu, Taiwan, R.O.C.}
\author{Yi Yang}
\email{yiyang@mail.nctu.edu.tw}
\affiliation{Department of Electrophysics, National Chiao-Tung University, Hsinchu, Taiwan, R.O.C.}
\author{}
\date{\today }

\begin{abstract}
We calculate a class of polarized fermion string scattering amplitudes (PFSSA)
at arbitrary mass levels. We discover that, in the hard scattering limit, the
functional forms of the non-vanishing PFSSA at each fixed mass level are
independent of the choices of spin polarizations. This result justifies and
extends Gross conjecture on high energy string scattering amplitudes to the
fermionic sector. In addition, this peculiar property of hard PFSSA is to be
compared with the usual spin polarization dependence of the hard polarized
fermion field theory scatterings.

\end{abstract}
\maketitle
\tableofcontents

%

%TCIMACRO{\TeXButton{equation number}{\setcounter{equation}{0}
%\renewcommand{\theequation}{\arabic{section}.\arabic{equation}}}}%
%BeginExpansion
\setcounter{equation}{0}
\renewcommand{\theequation}{\arabic{section}.\arabic{equation}}%
%EndExpansion

\section{Introduction}

One important characteristic of string scattering amplitudes (SSA) is its very
soft exponential fall-off behavior in the hard scattering limit. This behavior
is closely related to the existence of infinite linear relations among hard
SSA of different string states at each fixed mass level. Moreover, these
linear relations are so powerful that they can be used to solve all hard SSA
and express them in terms of one amplitude. This means that there is only one
hard SSA $f(E,\theta)$ at each fixed mass level which is very different from
the usual spin dependence of hard fermion field theory scatterings. This
important high energy symmetry of string theory was first conjectured by Gross
\cite{GM,Gross,GrossManes} and later corrected and proved by using the
decoupling of zero norm states \cite{ZNS1} in \cite{ChanLee1,ChanLee2,CHL,PRL,
CHLTY,susy}. For more details, see the recent review \cite{review}.

However, all calculations that have been done so far are only for boson SSA of
either the bosonic string theory \cite{ChanLee1,ChanLee2,CHL,PRL, CHLTY} or
the NS sector (both GSO even and odd) of the fermionic string theory
\cite{susy}. So it will be important and of interest to see whether one can
extend Gross conjecture to the R sector of the fermionic string theory.

Since it is a nontrivial task to construct the general massive fermion string
vertex operators, as the first step in this letter, we choose to calculate
polarized fermion string scattering amplitudes (PFSSA) at arbitrary mass
levels which involve the \textit{leading Regge trajectory} fermion string
state of the R sector ($\alpha^{\prime}\equiv\frac{1}{2}$) \cite{Osch}
\begin{equation}
\chi_{(m_{1}...m_{n-1}m_{n})}^{\alpha}i\partial X^{m_{1}}\cdots i\partial
X^{m_{n-1}}(i\partial X^{m_{n}}\delta_{\alpha}^{\gamma}-\frac{1}{8}%
\gamma_{\alpha\dot{\beta}}^{\mu}k_{\mu}\psi^{m_{n}}\gamma_{\nu}^{\dot{\beta
}\gamma}\psi^{\nu})S_{\gamma}e^{-\frac{\phi}{2}}e^{ikX}, \label{R}%
\end{equation}
in which the tensor-spinor wavefunction $\chi_{(m_{1}...m_{n-1}m_{n})}%
^{\alpha}$ satisfies the on-shell conditions%
\begin{equation}
k^{m_{i}}\chi_{(m_{1}...m_{n})}^{\alpha}=\eta^{m_{i}m_{j}}\chi_{(m_{1}%
...m_{n})}^{\alpha}=\chi_{(m_{1}...m_{n})}^{\alpha}\gamma_{\alpha\dot{\beta}%
}^{m_{i}}=0,M^{2}=2n, \label{con}%
\end{equation}
which include a $\gamma$ traceless condition. One of the reason for choosing
this leading Regge trajectory state is that the corresponding vertex operator
has been constructed in the literature \cite{Osch}. The construction was
mainly based on the complete construction of the first massive level states
for both NS and R sectors \cite{RRR}.

On the other hand, since Gross conjecture was shown to be valid for both GSO
even and odd states in the NS sector \cite{susy}, for simplicity in this
paper, we are going to ignore the GSO projection, and the other three string
states in the SSA will be chosen to be one massless fermion and two tachyon
states (GSO odd).

The state in Eq.(\ref{R}) is a combination of $(\alpha_{-1}^{i})^{n}\left\vert
\alpha\right\rangle _{R}$ and $(\alpha_{-1}^{i})^{n-1}(d_{-1}^{j})\left\vert
\bar{\alpha}\right\rangle _{R}$ (in the light-cone gauge language). For the
case of $n=1$ \cite{RRR}, for example, the vector-spinor $\chi_{\mu}^{\alpha}$
is a $10D$ Majorana spinor that forms an irreducible massive representation of
the Lorentz group. In the corresponding four dimensional case, the
vector-spinor $\chi_{\mu}^{\alpha}$ tranforms as the product of a four-vector
and a Dirac spinor, and satisfies the Rarita-Schwinger equations%
\begin{align}
(\gamma\cdot\partial+M)\chi_{\mu}  &  =0,\label{dd}\\
\gamma^{\mu}\chi_{\mu}  &  =0, \label{rr}%
\end{align}
which is the case of spin $s=\frac{3}{2}$ field equation of the more general
Bargmann-Wigner equation with spin $s\geq\frac{1}{2}$. Note that Eq.(\ref{rr})
is similar to the $\gamma$ traceless condition in Eq.(\ref{con}).

It was shown for the bosonic SSA that at each fixed mass level $M^{2}=2(N-1)$
only tensor states of the following form \cite{PRL, CHLTY}
\begin{equation}
\left\vert N,2m,q\right\rangle \equiv(\alpha_{-1}^{T})^{N-2m-2q}(\alpha
_{-1}^{L})^{2m}(\alpha_{-2}^{L})^{q}|0,k\rangle\label{Nmq}%
\end{equation}
are of leading order in energy in the hard scattering limit. In Eq.(\ref{Nmq}%
), $e^{P}=\frac{1}{M}(E,\mathrm{k},0)=\frac{k_{2}}{M_{2}}$ the momentum
polarization, $e^{L}=\frac{1}{M}(\mathrm{k},E,0)$ the longitudinal
polarization and $e^{T}=(0,0,1)$ the transverse polarization are the three
polarizations on the scattering plane \cite{ChanLee1,ChanLee2}. In the hard
scattering limit, one can identify $e^{P}=$ $e^{L}$ \cite{ChanLee1,ChanLee2}.
It was remarkable to discover that all the hard bosonic SSA at each fixed mass
level share the same functional forms with the following ratios \cite{PRL,
CHLTY}%
\begin{equation}
\frac{T^{(N,2m,q)}}{T^{(N,0,0)}}=\left(  -\frac{1}{M}\right)  ^{2m+q}\left(
\frac{1}{2}\right)  ^{m+q}(2m-1)!!.\label{04}%
\end{equation}
Thus there is only one hard SSA $T^{(N,0,0)}$ at each fixed mass level. For
the leading Regge trajectory states we are considering in this paper, we set
$q=0$.

In this paper we will be mainly concerning with the spinor polarizations in
the SSA calculation. So, for simplicity, we will be writing%
\begin{equation}
\chi_{(m_{1}...m_{n})}^{\alpha}\sim\epsilon_{m_{1}}...\epsilon_{m_{n}%
}u^{\alpha}%
\end{equation}
where $u^{\alpha}$ satisfies the $10D$ Dirac type equation in Eq.(\ref{dd}).
For the leading hard SSA of the Regge trajectory states, one can choose to put
all the tensor polarizations $\epsilon_{m_{1}}=\cdots=\epsilon_{m_{n}%
}=\epsilon_{T}$.%

%TCIMACRO{\TeXButton{equation number}{\setcounter{equation}{0}
%\renewcommand{\theequation}{\arabic{section}.\arabic{equation}}}}%
%BeginExpansion
\setcounter{equation}{0}
\renewcommand{\theequation}{\arabic{section}.\arabic{equation}}%
%EndExpansion

\section{Polarized fermion string scattering amplitudes (PFSSA)}

In this section, we will calculate the PFSSA with the following four vertex
operators in the $10D$ open superstring theory: \bigskip the massless spinor%
\begin{equation}
V=u^{\alpha}S_{\alpha}e^{-\frac{\phi}{2}}e^{ikX},
\end{equation}
the massive spinor $(\alpha^{\prime}=\frac{1}{2})$%
\begin{align}
V  &  =\left(  i\epsilon_{1}\partial X\right)  \cdots\left(  i\epsilon
_{n}\partial X\right)  u^{\alpha}S_{\alpha}e^{-\frac{\phi}{2}}e^{ikX}\\
&  -\frac{1}{8}\left(  i\epsilon_{1}\partial X\right)  \cdots\left(
i\epsilon_{n-1}\partial X\right)  u^{\alpha}\gamma_{\alpha\dot{\beta}}^{\mu
}k_{\mu}\psi_{n}\gamma_{\nu}^{\dot{\beta}\gamma}\psi^{\nu}S_{\gamma}%
e^{-\frac{\phi}{2}}e^{ikX}\nonumber
\end{align}
and two tachyons%
\begin{equation}
V=e^{-\phi}e^{ikX}\text{ \ \ }(-1\text{ }ghost),
\end{equation}%
\begin{equation}
V=\left(  k\cdot\psi\right)  e^{ikX}\text{ \ \ }(0\text{ }ghost).
\end{equation}
In the above, we have chosen the total ghost charges sum up to $-2$. The
correlators of the worldsheet boson $X^{\mu}$, worldsheet fermion $\psi^{\mu}%
$, spin field $S_{A}$ and ghost field $\phi$ are%
\begin{align}
\left\langle X^{\mu}\left(  z_{1}\right)  X^{\nu}\left(  z_{2}\right)
\right\rangle  &  =-\eta^{\mu\nu}\ln\left\vert z_{12}\right\vert \\
\left\langle \psi^{\mu}\left(  z_{1}\right)  S_{A}\left(  z_{2}\right)
S_{B}\left(  z_{3}\right)  \right\rangle  &  =\frac{\left(  \Gamma^{\mu
}C\right)  _{AB}}{\sqrt{2}z_{12}^{\frac{1}{2}}z_{13}^{\frac{1}{2}}%
z_{23}^{\frac{3}{2}}}\label{ss}\\
\left\langle \psi^{\mu}\left(  z_{1}\right)  \psi^{\nu}\left(  z_{2}\right)
\right\rangle  &  =\frac{\eta^{\mu\nu}}{z_{12}}\\
\left\langle \phi\left(  z_{1}\right)  \phi\left(  z_{2}\right)
\right\rangle  &  =-\ln\left\vert z_{12}\right\vert
\end{align}
where $z_{ij}=z_{i}-z_{j}$, $\Gamma^{\mu}$ are $10D$ Dirac matrices calculated
in Eq.(\ref{gamma}) and $C$ matrix calculated in Eq.(\ref{cc}).

The PFSSA we want to calculate can be written as%
\begin{equation}
A=A_{1}+A_{2} \label{00}%
\end{equation}
where%
\begin{align}
A_{1}  &  =\int dz_{2}\left\vert z_{13}z_{14}z_{34}\right\vert \left\langle
\begin{array}
[c]{c}%
u_{1}^{\alpha_{1}}S_{1\alpha_{1}}e^{\frac{-\phi_{1}}{2}}e^{ik_{1}X_{1}}\left(
i\epsilon_{1}\partial X_{2}\right)  \cdots\left(  i\epsilon_{n}\partial
X_{2}\right)  u_{2}^{\alpha_{2}}S_{2\alpha_{2}}e^{\frac{-\phi_{2}}{2}%
}e^{ik_{2}X_{2}}\\
e^{-\phi_{3}}e^{ik_{3}X_{3}}\left(  k_{4}\cdot\psi_{4}\right)  e^{ik_{4}X_{4}}%
\end{array}
\right\rangle ,\\
A_{2}  &  =-\frac{1}{8}\int dz_{2}\left\vert z_{13}z_{14}z_{34}\right\vert
\left\langle
\begin{array}
[c]{c}%
u_{1}^{\alpha_{1}}S_{1\alpha_{1}}e^{\frac{-\phi_{1}}{2}}e^{ik_{1}X_{1}}\\
\left(  i\epsilon_{1}\partial X_{2}\right)  \cdots\left(  i\epsilon
_{n-1}\partial X_{2}\right)  u_{2}^{\alpha_{2}}\gamma_{\alpha_{2}\dot{\beta}%
}^{\mu}k_{2\mu}\psi_{2n}\gamma_{\nu}^{\dot{\beta}\gamma}\psi_{2}^{\nu
}S_{2\gamma}e^{-\frac{\phi_{2}}{2}}e^{ik_{2}X_{2}}\\
e^{-\phi_{3}}e^{ik_{3}X_{3}}\left(  k_{4}\cdot\psi_{4}\right)  e^{ik_{4}X_{4}}%
\end{array}
\right\rangle . \label{a2}%
\end{align}

Let's calculate $A_{1}$ first, whose correlator can be written as%

\begin{equation}
u_{1}^{\alpha_{1}}u_{2}^{\alpha_{2}}k_{4\mu}\left\langle S_{1\alpha_{1}%
}S_{2\alpha_{2}}\psi_{4}^{\mu}\right\rangle \left\langle e^{\frac{-\phi_{1}%
}{2}}e^{\frac{-\phi_{2}}{2}}e^{-\phi_{3}}\right\rangle \left\langle
e^{ik_{1}X_{1}}\left(  i\epsilon_{1}\partial X_{2}\right)  \cdots\left(
i\epsilon_{n}\partial X_{2}\right)  e^{ik_{2}X_{2}}e^{ik_{3}X_{3}}%
e^{ik_{4}X_{4}}\right\rangle . \label{3cor}%
\end{equation}
The first correlators in Eq.(\ref{3cor}) was calculated in Eq.(\ref{ss})
\cite{Osch}, and the other two can be calculated to be%
\begin{equation}
\left\langle e^{\frac{-\phi_{1}}{2}}e^{\frac{-\phi_{2}}{2}}e^{-\phi_{3}%
}\right\rangle =\left\vert z_{12}\right\vert ^{\frac{-1}{4}}\left\vert
z_{13}\right\vert ^{\frac{-1}{2}}\left\vert z_{23}\right\vert ^{\frac{-1}{2}}
\label{gho}%
\end{equation}
and%
\begin{align}
&  \left\langle e^{ik_{1}X_{1}}\left(  i\epsilon_{1}\partial X_{2}\right)
\cdots\left(  i\epsilon_{n}\partial X_{2}\right)  e^{ik_{2}X_{2}}%
e^{ik_{3}X_{3}}e^{ik_{4}X_{4}}\right\rangle \nonumber\\
&  =\left\vert z_{12}\right\vert ^{k_{1}\cdot k_{2}}\left\vert z_{13}%
\right\vert ^{k_{1}\cdot k_{3}}\left\vert z_{14}\right\vert ^{k_{1}\cdot
k_{4}}\left\vert z_{23}\right\vert ^{k_{2}\cdot k_{3}}\left\vert
z_{24}\right\vert ^{k_{2}\cdot k_{4}}\left\vert z_{34}\right\vert ^{k_{3}\cdot
k_{4}}\nonumber\\
&  \times\left(  \frac{k_{1}\cdot\epsilon_{1}}{z_{21}}+\frac{k_{3}%
\cdot\epsilon_{1}}{z_{23}}+\frac{k_{4}\cdot\epsilon_{1}}{z_{24}}\right)
\cdots\left(  \frac{k_{1}\cdot\epsilon_{n}}{z_{21}}+\frac{k_{3}\cdot
\epsilon_{n}}{z_{23}}+\frac{k_{4}\cdot\epsilon_{n}}{z_{24}}\right)  .
\label{old}%
\end{align}

For the $s-t$ channel amplitude, we take $z_{1}=0,z_{3}=1,z_{4}\rightarrow
\infty$ $\left(  0\leq z_{2}\leq1\right)  $ and, for simplicity, set all
$\epsilon_{1}=\cdots=\epsilon_{n}=\epsilon$, we get%
\begin{equation}
A_{1}=\frac{1}{\sqrt{2}}u_{1}^{\alpha_{1}}\left(  \Gamma^{\mu}C\right)
_{\alpha_{1}\alpha_{2}}u_{2}^{\alpha_{2}}k_{4\mu}\left(  -1\right)  ^{\frac
{3}{4}}\int_{0}^{1}dz_{2}z_{2}^{k_{1}\cdot k_{2}-1}\left(  1-z_{2}\right)
^{k_{2}\cdot k_{3}-\frac{1}{2}}\left(  \frac{k_{1}\cdot\epsilon}{z_{2}}%
-\frac{k_{3}\cdot\epsilon}{1-z_{2}}\right)  ^{n}.
\end{equation}

Finally the integration in $A_{1}$ can be performed \cite{LLY2} and we obtain%
\begin{align}
A_{1}  &  =\frac{1}{\sqrt{2}}u_{1}^{\alpha_{1}}\left(  \Gamma^{\mu}C\right)
_{\alpha_{1}\alpha_{2}}u_{2}^{\alpha_{2}}k_{4\mu}\times\label{11}\\
&  \left(  -1\right)  ^{\frac{3}{4}}\left(  -k_{3}\cdot\epsilon\right)
^{n}B\left(  \frac{-s}{2}+n;\frac{-t}{2}\right)  F_{D}^{\left(  1\right)
}\left(  \frac{-t}{2};-n;\frac{s}{2}-n+1;\frac{-k_{1}\cdot\epsilon}{k_{3}%
\cdot\epsilon}\right)  \label{111}%
\end{align}
where $s=-\left(  k_{1}+k_{2}\right)  ^{2}$ and $t=-\left(  k_{2}%
+k_{3}\right)  ^{2}$ are the Mandelstam variables, and $F_{D}^{\left(
1\right)  }$ is the Lauricella function $(K=1)$ \cite{Appell}%
\begin{align}
&  F_{D}^{(K)}\left(  \alpha;\beta_{1},...,\beta_{K};\gamma;x_{1}%
,...,x_{K}\right) \nonumber\\
&  =\frac{\Gamma(\gamma)}{\Gamma(\alpha)\Gamma(\gamma-\alpha)}\int_{0}%
^{1}dt\,t^{\alpha-1}(1-t)^{\gamma-\alpha-1}\cdot(1-x_{1}t)^{-\beta_{1}%
}(1-x_{2}t)^{-\beta_{2}}...(1-x_{K}t)^{-\beta_{K}}. \label{Kam}%
\end{align}

Similar techanique can be used to calculate $A_{2}$ whose correlator can be
written as%

\begin{align}
&  u_{1}^{\alpha_{1}}u_{2}^{\alpha_{2}}\epsilon_{n}^{\lambda}k_{4\nu
}\left\langle S_{1\alpha_{1}}\gamma_{\alpha_{2}\dot{\beta}}^{\mu}k_{2\mu}%
\psi_{2\lambda}\gamma_{\rho}^{\dot{\beta}\gamma}\psi_{2}^{\rho}S_{2\gamma}%
\psi_{4}^{\nu}\right\rangle \left\langle e^{\frac{-\phi_{1}}{2}}e^{\frac
{-\phi_{2}}{2}}e^{-\phi_{3}}\right\rangle \nonumber\\
&  \times\left\langle e^{ik_{1}X_{1}}\left(  i\epsilon_{1}\partial
X_{2}\right)  \cdots\left(  i\epsilon_{n-1}\partial X_{2}\right)
e^{ik_{2}X_{2}}e^{ik_{3}X_{3}}e^{ik_{4}X_{4}}\right\rangle . \label{4cor}%
\end{align}
The second and the third correlators in Eq.(\ref{4cor}) were calculated in
Eq.(\ref{gho}) and Eq.(\ref{old}) respectively. The first correlator can be
written as%
\begin{equation}
\left\langle S_{1\alpha_{1}}\gamma_{\alpha_{2}\dot{\beta}}^{\mu}k_{2\mu}%
\psi_{2\lambda}\gamma_{\rho}^{\dot{\beta}\gamma}\psi_{2}^{\rho}S_{2\gamma}%
\psi_{4}^{\nu}\right\rangle =\left\langle S_{1\alpha_{1}}\gamma_{\alpha
_{2}\dot{\beta}}^{\mu}k_{2\mu}K_{\lambda}^{\dot{\beta}}\psi_{4}^{\nu
}\right\rangle \label{kk}%
\end{equation}
where the composite operators $K_{\lambda}^{\dot{\beta}}$ was defined to be
\cite{Osch}%
\begin{equation}
K_{\lambda}^{\dot{\beta}}=\psi_{2\lambda}\gamma_{\rho}^{\dot{\beta}\gamma}%
\psi_{2}^{\rho}S_{2\gamma}.
\end{equation}

The correlation functions containing spin fields $S_{\alpha}$ and the
composite operators $K_{\lambda}^{\dot{\beta}}$ can be found in \cite{Osch}.
The computation of correlation functions with $K_{\lambda}^{\dot{\beta}}$ got
simplified due to the $\gamma$ traceless condition in Eq.(\ref{con}). The
correlator in Eq.(\ref{kk}) can then be calculated to be%
\begin{equation}
\left\langle S_{1\alpha_{1}}\gamma_{\alpha_{2}\dot{\beta}}^{\mu}k_{2\mu
}K_{\lambda}^{\dot{\beta}}\psi_{4}^{\nu}\right\rangle =\frac{\left(
10-2\right)  z_{41}^{\frac{1}{2}}}{\sqrt{2}z_{42}^{\frac{3}{2}}z_{12}%
^{\frac{10}{8}+\frac{1}{2}}}\gamma_{\alpha_{2}\dot{\beta}}^{\mu}k_{2\mu}%
\eta_{\lambda}^{\nu}C_{\alpha_{1}}^{\dot{\beta}}=\frac{8z_{41}^{\frac{1}{2}%
}\gamma_{\alpha_{2}\dot{\beta}}^{\mu}k_{2\mu}C_{\alpha_{1}}^{\dot{\beta}}%
\eta_{\lambda}^{\nu}}{\sqrt{2}z_{42}^{\frac{3}{2}}z_{12}^{\frac{7}{4}}}.
\label{kkk}%
\end{equation}
By using Eq.(\ref{gho}), Eq.(\ref{old}) and Eq.(\ref{kkk}), we get
\begin{align}
A_{2}  &  =-\frac{1}{\sqrt{2}}u_{1}^{\alpha_{1}}u_{2}^{\alpha_{2}}%
\gamma_{\alpha_{2}\dot{\beta}}^{\mu}C_{\alpha_{1}}^{\dot{\beta}}k_{2\mu
}\left(  k_{4}\cdot\epsilon_{n}\right) \nonumber\\
&  \times\int dz_{2}\frac{\left\vert z_{12}\right\vert ^{k_{1}\cdot
k_{2}-\frac{1}{4}}}{z_{12}^{\frac{7}{4}}}\left\vert z_{13}\right\vert
^{k_{1}\cdot k_{3}+\frac{1}{2}}\left\vert z_{14}\right\vert ^{k_{1}\cdot
k_{4}+1}z_{41}^{\frac{1}{2}}\left\vert z_{23}\right\vert ^{k_{2}\cdot
k_{3}-\frac{1}{2}}\frac{\left\vert z_{24}\right\vert ^{k_{2}\cdot k_{4}}%
}{z_{42}^{\frac{3}{2}}}\left\vert z_{34}\right\vert ^{k_{3}\cdot k_{4}%
+1}\nonumber\\
&  \times\left(  \frac{k_{1}\cdot\epsilon_{1}}{z_{21}}+\frac{k_{3}%
\cdot\epsilon_{1}}{z_{23}}+\frac{k_{4}\cdot\epsilon_{1}}{z_{24}}\right)
\cdots\left(  \frac{k_{1}\cdot\epsilon_{n-1}}{z_{21}}+\frac{k_{3}\cdot
\epsilon_{n-1}}{z_{23}}+\frac{k_{4}\cdot\epsilon_{n-1}}{z_{24}}\right)  .
\end{align}

For the $s-t$ channel amplitude, we take $z_{1}=0,z_{3}=1,z_{4}\rightarrow
\infty$ $\left(  0\leq z_{2}\leq1\right)  $ and, for simplicity, set all
$\epsilon_{1}=\cdots=\epsilon_{n}=\epsilon$ as before, we get%
\begin{equation}
A_{2}=-\frac{1}{\sqrt{2}}u_{1}^{\alpha_{1}}u_{2}^{\alpha_{2}}\gamma
_{\alpha_{2}\dot{\beta}}^{\mu}C_{\alpha_{1}}^{\dot{\beta}}k_{2\mu}\left(
k_{4}\cdot\epsilon\right)  \left(  -1\right)  ^{\frac{3}{4}}\int_{0}^{1}%
dz_{2}z_{2}^{\frac{-s}{2}+n-2}\left(  1-z_{2}\right)  ^{\frac{-t}{2}%
+n-1}\left(  \frac{k_{1}\cdot\epsilon}{z_{2}}-\frac{k_{3}\cdot\epsilon
}{1-z_{2}}\right)  ^{n-1}.
\end{equation}

Finally the integration in $A_{2}$ can be performed \cite{LLY2} and we obtain%
\begin{align}
A_{2}  &  =\frac{-1}{\sqrt{2}}u_{1}^{\alpha_{1}}\left(  \Gamma^{\mu}C\right)
_{\alpha_{1}\alpha_{2}}u_{2}^{\alpha_{2}}k_{2\mu}\times\label{2222}\\
&  \left(  -1\right)  ^{\frac{3}{4}}\left(  k_{4}\cdot\epsilon\right)  \left(
-k_{3}\cdot\epsilon\right)  ^{n-1}B\left(  \frac{-s}{2}+n-1;\frac{-t}%
{2}+1\right)  F_{D}^{\left(  1\right)  }\left(  \frac{-t}{2}+1;-n+1;\frac
{s}{2}-n+2;\frac{-k_{1}\cdot\epsilon}{k_{3}\cdot\epsilon}\right)  .
\label{222}%
\end{align}
This completes the calculation of the PFSSA.

\section{Hard scattering limit}

In this section, we will calculate the hard scattering limit of the PFSSA we
obtained in the previous section. We will concentrate on the spinor
polarizations and ignore the parts of the tensor polarizations. To do so we
need to solve $10D$ Dirac equation and calculate explicitly the two factors in
Eq.(\ref{11}) and Eq.(\ref{2222})
\begin{align}
&  u_{1}^{\alpha_{1}}\left(  \Gamma^{\mu}C\right)  _{\alpha_{1}\alpha_{2}%
}u_{2}^{\alpha_{2}}k_{4\mu},\label{33}\\
&  u_{1}^{\alpha_{1}}\left(  \Gamma^{\mu}C\right)  _{\alpha_{1}\alpha_{2}%
}u_{2}^{\alpha_{2}}k_{2\mu}. \label{44}%
\end{align}

We will follow the definition in \cite{Polchin} to calculate the $10D$ Dirac
matrices. The ground states of the R sector are degenerate and can be labeled
by%
\begin{equation}
\mathbf{s=(}s_{0},s_{1},s_{2},s_{3},s_{4})
\end{equation}
where each of the $s_{a}$ is $\pm\frac{1}{2}$ in the $\mathbf{s}$ basis. To
simplify the notation, we will ignore the factor $\frac{1}{2}$ in the rest of
the paper. There are $2^{\frac{10}{2}}=32$ components of a $10D$ Dirac spinor.
The $10D$ Dirac matrices can be calculated iteratively starting in $d=2$,
where%
\begin{equation}
\Gamma^{0}=i\sigma^{2},\Gamma^{1}=\sigma^{1}.
\end{equation}
Then in $d=2k+2$,%
\begin{align}
\Gamma^{\mu}  &  =\gamma^{\mu}\otimes(-\sigma^{3}),\mu=0,\cdots,d-3,\\
\Gamma^{d-2}  &  =I\otimes\sigma^{1},\text{ }\Gamma^{d-1}=I\otimes\sigma^{2}%
\end{align}
where $\gamma^{\mu}$ is the $2^{k}\times$ $2^{k}$ Dirac matrices in $d-2$
dimensions and $I$ is the $2^{k}\times$ $2^{k}$ identity matrix. We list all
the $10D$ Dirac matrices calculated in the following%
\begin{align}
\Gamma^{0}  &  =i\sigma^{2}\otimes\sigma^{3}\otimes\sigma^{3}\otimes\sigma
^{3}\otimes\sigma^{3},\nonumber\\
\Gamma^{1}  &  =\sigma^{1}\otimes\sigma^{3}\otimes\sigma^{3}\otimes\sigma
^{3}\otimes\sigma^{3},\nonumber\\
\Gamma^{2}  &  =-I_{2}\otimes\sigma^{1}\otimes\sigma^{3}\otimes\sigma
^{3}\otimes\sigma^{3},\nonumber\\
\Gamma^{3}  &  =-I_{2}\otimes\sigma^{2}\otimes\sigma^{3}\otimes\sigma
^{3}\otimes\sigma^{3},\nonumber\\
\Gamma^{4}  &  =I_{2}\otimes I_{2}\otimes\sigma^{1}\otimes\sigma^{3}%
\otimes\sigma^{3},\nonumber\\
\Gamma^{5}  &  =I_{2}\otimes I_{2}\otimes\sigma^{2}\otimes\sigma^{3}%
\otimes\sigma^{3},\nonumber\\
\Gamma^{6}  &  =-I_{2}\otimes I_{2}\otimes I_{2}\otimes\sigma^{1}\otimes
\sigma^{3},\nonumber\\
\Gamma^{7}  &  =-I_{2}\otimes I_{2}\otimes I_{2}\otimes\sigma^{2}\otimes
\sigma^{3},\nonumber\\
\Gamma^{8}  &  =I_{2}\otimes I_{2}\otimes I_{2}\otimes I_{2}\otimes\sigma
^{1},\nonumber\\
\Gamma^{9}  &  =I_{2}\otimes I_{2}\otimes I_{2}\otimes I_{2}\otimes\sigma^{2}.
\label{gamma}%
\end{align}

\bigskip We begin with the calculation of $C$ matrix in Eq.(\ref{33}) and
Eq.(\ref{44}), which is defined to be%
\begin{equation}
C=B_{1}\Gamma^{0}\text{ }%
\end{equation}
where%
\begin{align}
B_{1}  &  =\Gamma^{3}\Gamma^{5}\Gamma^{7}\Gamma^{9}\nonumber\\
&  =-I_{2}\otimes\sigma^{2}\otimes\sigma^{1}\otimes\sigma^{2}\otimes\sigma
^{1}.
\end{align}
So we have%
\begin{equation}
C=B_{1}\Gamma^{0}=-i\sigma^{2}\otimes\sigma^{1}\otimes\sigma^{2}\otimes
\sigma^{1}\otimes\sigma^{2} \label{cc}%
\end{equation}
and%
\begin{align}
\Gamma^{0}C  &  =I_{2}\otimes\sigma^{2}\otimes\sigma^{1}\otimes\sigma
^{2}\otimes\sigma^{1},\label{r1}\\
\Gamma^{1}C  &  =\sigma^{3}\otimes\sigma^{2}\otimes\sigma^{1}\otimes\sigma
^{2}\otimes\sigma^{1},\label{r2}\\
\Gamma^{2}C  &  =\sigma^{2}\otimes I_{2}\otimes\sigma^{1}\otimes\sigma
^{2}\otimes\sigma^{1}, \label{r3}%
\end{align}

The next step is to solve $10D$ Dirac equation%
\begin{equation}
\left(  ik\cdot\Gamma+M\right)  u=0,
\end{equation}
and calculate explicitly the spinors $u_{1}$ and $u_{2}$ in Eq.(\ref{33}) and
Eq.(\ref{44}). In the CM frame, we have the kinematics%
\begin{align*}
k_{1}  &  =\left(  +\sqrt{p^{2}+M_{1}^{2}},-p,0\right)  ,\\
k_{2}  &  =\left(  +\sqrt{p^{2}+M_{2}^{2}},+p,0\right)  ,\\
k_{3}  &  =\left(  -\sqrt{q^{2}+M_{3}^{2}},-q\cos\theta,-q\sin\theta\right)
,\\
k_{4}  &  =\left(  -\sqrt{q^{2}+M_{4}^{2}},+q\cos\theta,+q\sin\theta\right)  .
\end{align*}
For our case here, $u_{1}$ is a massless spinor, so we have%
\begin{equation}
k_{1}^{\mu}=\left(  +p,-p,0\right)  ,
\end{equation}%
\begin{equation}
ik_{1}\cdot\Gamma u_{1}=0.
\end{equation}
The $10D$ Dirac equation can be calculated to be%
\begin{align}
&  \left[  i\left(  -p\right)  \Gamma^{0}+i\left(  -p\right)  \Gamma
^{1}\right]  u_{1}\nonumber\\
&  =-ip\left[  i\sigma^{2}\otimes\sigma^{3}\otimes\sigma^{3}\otimes\sigma
^{3}\otimes\sigma^{3}+\sigma^{1}\otimes\sigma^{3}\otimes\sigma^{3}%
\otimes\sigma^{3}\otimes\sigma^{3}\right]  u_{1}\nonumber\\
&  =\left(  i\sigma^{2}+\sigma^{1}\right)  \otimes\sigma^{3}\otimes\sigma
^{3}\otimes\sigma^{3}\otimes\sigma^{3}u_{1}=0,
\end{align}
or%
\begin{equation}%
\begin{pmatrix}
0 & 2\\
0 & 0
\end{pmatrix}
\otimes\sigma^{3}\otimes\sigma^{3}\otimes\sigma^{3}\otimes\sigma^{3}u_{1}=0,
\end{equation}
which can be solved to be%
\begin{equation}
u_{1}=%
\begin{pmatrix}
1\\
0
\end{pmatrix}
\otimes\left\vert \pm;\pm;\pm;\pm\right\rangle
\end{equation}
where
\begin{equation}
\left\vert +\right\rangle =%
\begin{pmatrix}
1\\
0
\end{pmatrix}
,\left\vert -\right\rangle =%
\begin{pmatrix}
0\\
1
\end{pmatrix}
.
\end{equation}

For the massive $u_{2}$ we have%
\begin{equation}
\left(  ik_{2\mu}\Gamma^{\mu}+M_{2}\right)  u_{2}=0,
\end{equation}%
\begin{equation}
k_{2}^{\mu}=\left(  +E_{2},+p,0\right)  =\left(  +\sqrt{p^{2}+M_{2}^{2}%
},+p,0\right)  .
\end{equation}
The $10D$ Dirac equation can be calculated to be%
\begin{align}
&  \left[  i\left(  -E_{2}\right)  \Gamma^{0}+ip\Gamma^{1}+M_{2}\right]
u_{2}\nonumber\\
&  =\left[  i\left(  -E_{2}\right)  i\sigma^{2}\otimes\sigma^{3}\otimes
\sigma^{3}\otimes\sigma^{3}\otimes\sigma^{3}+ip\sigma^{1}\otimes\sigma
^{3}\otimes\sigma^{3}\otimes\sigma^{3}\otimes\sigma^{3}+M_{2}\right]
u_{2}\nonumber\\
&  =\left[  \left(  -iE_{2}i\sigma^{2}+ip\sigma^{1}\right)  \otimes\sigma
^{3}\otimes\sigma^{3}\otimes\sigma^{3}\otimes\sigma^{3}+M_{2}I_{2}\otimes
I_{2}\otimes I_{2}\otimes I_{2}\otimes I_{2}\right]  u_{2}=0. \label{d2}%
\end{align}

Let's first assume%
\begin{equation}
u_{2}=U_{2}\otimes\left\vert \text{even \# of }"-"\right\rangle \label{u2}%
\end{equation}
where $U_{2}$ is a $2$-spinor. If we put Eq.(\ref{u2}) into Eq.(\ref{d2}), we get%

\begin{equation}%
\begin{pmatrix}
M_{2} & -iE_{2}+ip\\
iE_{2}+ip & M_{2}%
\end{pmatrix}
U_{2}=0,
\end{equation}
which can be solved to be%
\begin{equation}
U_{2}=%
\begin{pmatrix}
i\sqrt{\frac{E_{2}-p}{2E_{2}}}\\
\frac{M_{2}}{\sqrt{2E_{2}\left(  E_{2}-p\right)  }}%
\end{pmatrix}
.
\end{equation}
So the first class of solutions of $u_{2}$ is%
\begin{equation}
u_{2}^{(1)}=%
\begin{pmatrix}
i\sqrt{\frac{E_{2}-p}{2E_{2}}}\\
\frac{M_{2}}{\sqrt{2E_{2}\left(  E_{2}-p\right)  }}%
\end{pmatrix}
\otimes\left\vert \text{even \# of }"-"\right\rangle . \label{low}%
\end{equation}
Alternatively, we can assume%
\begin{equation}
u_{2}=U_{2}\otimes\left\vert \text{odd \# of }"-"\right\rangle .
\end{equation}
For this case, Dirac equation reduces to%

\begin{equation}%
\begin{pmatrix}
M_{2} & iE_{2}-ip\\
-iE_{2}-ip & M_{2}%
\end{pmatrix}
U_{2}=0,
\end{equation}
and we get the second class of solutions%
\begin{equation}
u_{2}^{(2)}=%
\begin{pmatrix}
i\sqrt{\frac{E_{2}-p}{2E_{2}}}\\
\frac{-M_{2}}{\sqrt{2E_{2}\left(  E_{2}-p\right)  }}%
\end{pmatrix}
\otimes\left\vert \text{odd \# of }"-"\right\rangle . \label{high}%
\end{equation}

We are now ready to calculate the vector components of $u_{1}\Gamma^{\mu
}Cu_{2}$ in Eq.(\ref{33}) and Eq.(\ref{44}) which are to be contracted with
$k_{4}$ and $k_{2}$. One needs only calculate the first three components of
the vector.

On the other hand, it is crucial to note that the last three components of
$\Gamma^{0}C$, $\Gamma^{1}C$ and $\Gamma^{2}C$ in Eq.(\ref{r1}), Eq.(\ref{r2})
and Eq.(\ref{r3}) are all off-diagonal matrices. In order to get non-vanishing
amplitudes, one is forced to choose different spin sign factors for each of
the last three spin components of $u_{1}$ and $u_{2}$. We will see that the
choice of $u_{2}^{(2)}$ in Eq.(\ref{high}) give leading order amplitudes in
the hard scattering limit, while the choice of $u_{2}^{(1)}$ in Eq.(\ref{low})
give subleading order amplitudes.

\bigskip For the first case, as an example, we choose $u_{1}$ as
\begin{equation}
u_{1}=%
\begin{pmatrix}
1\\
0
\end{pmatrix}
\otimes\left\vert +;+;+;+\right\rangle \label{u1}%
\end{equation}
and $u_{2}$ as%
\begin{equation}
u_{2}^{(1)}=%
\begin{pmatrix}
i\sqrt{\frac{E_{2}-p}{2E_{2}}}\\
\frac{M_{2}}{\sqrt{2E_{2}\left(  E_{2}-p\right)  }}%
\end{pmatrix}
\otimes\left\vert \text{even \# of }"-"\right\rangle =%
\begin{pmatrix}
i\sqrt{\frac{E_{2}-p}{2E_{2}}}\\
\frac{M_{2}}{\sqrt{2E_{2}\left(  E_{2}-p\right)  }}%
\end{pmatrix}
\otimes\left\vert -;-;-;-\right\rangle .
\end{equation}
The first three component of $u_{1}\Gamma^{\mu}Cu_{2}^{(1)}$ can be calculated
to be%
\begin{equation}
u_{1}\Gamma^{0}Cu_{2}^{(1)}=-i\sqrt{\frac{E_{2}-p}{2E_{2}}},
\end{equation}%
\begin{equation}
u_{1}\Gamma^{1}Cu_{2}^{(1)}=-i\sqrt{\frac{E_{2}-p}{2E_{2}}}%
\end{equation}
and%
\begin{equation}
u_{1}\Gamma^{2}Cu_{2}^{(1)}=0.
\end{equation}
So we have in Eq.(\ref{33})%
\begin{equation}
u_{1}\Gamma^{\mu}Cu_{2}^{(1)}k_{4\mu}=-i\sqrt{\frac{E_{2}-p}{2E_{2}}}%
(E_{4}+q\cos\theta) \label{hard1}%
\end{equation}
and in Eq.(\ref{44})%
\begin{equation}
u_{1}\Gamma^{\mu}Cu_{2}^{(1)}k_{2\mu}=-i\sqrt{\frac{E_{2}-p}{2E_{2}}}%
(-E_{2}+p). \label{hard2}%
\end{equation}

For the second case, as an example, we choose $u_{1}$ as in Eq.(\ref{u1}) and
$u_{2}$ as%
\begin{equation}
u_{2}^{(2)}=%
\begin{pmatrix}
i\sqrt{\frac{E_{2}-p}{2E_{2}}}\\
\frac{-M_{2}}{\sqrt{2E_{2}\left(  E_{2}-p\right)  }}%
\end{pmatrix}
\otimes\left\vert \text{odd \# of }"-"\right\rangle =%
\begin{pmatrix}
i\sqrt{\frac{E_{2}-p}{2E_{2}}}\\
\frac{-M_{2}}{\sqrt{2E_{2}\left(  E_{2}-p\right)  }}%
\end{pmatrix}
\otimes\left\vert +;-;-;-\right\rangle .
\end{equation}
The first three component of $u_{1}\Gamma^{\mu}Cu_{2}^{(2)}$ can be calculated
to be%
\begin{equation}
u_{1}\Gamma^{0}Cu_{2}^{(2)}=0,
\end{equation}%
\begin{equation}
u_{1}\Gamma^{1}Cu_{2}^{(2)}=0
\end{equation}
and%
\begin{equation}
u_{1}\Gamma^{2}Cu_{2}^{(2)}=\frac{M_{2}}{\sqrt{2E_{2}\left(  E_{2}-p\right)
}}.
\end{equation}
So we have in Eq.(\ref{33})%
\begin{equation}
u_{1}\Gamma^{\mu}Cu_{2}^{(2)}k_{4\mu}=\frac{M_{2}}{\sqrt{2E_{2}\left(
E_{2}-p\right)  }}q\sin\theta\label{hard3}%
\end{equation}
and in Eq.(\ref{44})%
\begin{equation}
u_{1}\Gamma^{\mu}Cu_{2}^{(2)}k_{2\mu}=0. \label{hard4}%
\end{equation}

\bigskip In the hard scattering limit, the energy order of Eq.(\ref{111}) and
Eq.(\ref{222}) are the same. To calculate the leading order amplitudes in
Eq.(\ref{00}), we need the results calculated in Eq.(\ref{hard1}),
Eq.(\ref{hard2}), Eq.(\ref{hard3}) and Eq.(\ref{hard4}). We note that
\begin{equation}
p=\sqrt{E_{2}^{2}-M_{2}^{2}}=E_{2}\left(  1-\frac{M_{2}^{2}}{2E_{2}^{2}%
}+\cdots\right)  ,
\end{equation}
so one gets in the hard scattering limit%
\begin{equation}
\frac{M_{2}}{\sqrt{2E_{2}\left(  E_{2}-p\right)  }}\rightarrow1,
\end{equation}%
\begin{equation}
\sqrt{\frac{E_{2}-p}{2E_{2}}}\rightarrow0.
\end{equation}
Finally the only leading order amplitude in the hard scattering limit is%
\begin{equation}
u_{1}\Gamma^{\mu}Cu_{2}^{(2)}k_{4\mu}=E\sin\theta.
\end{equation}

We conclude that for the choice of Eq.(\ref{high}), $A_{1}$ in Eq.(\ref{00})
gives the leading order amplitudes in the hard scattering limit. One can count
the number of leading order amplitudes. There are $2^{4}=16$ choices of spin
polarizations for $u_{1}$. Once the polarization of $u_{1}$ is fixed, each of
the last three spin signs of $u_{2}$ are fixed to be of the different sign
with $u_{1}$, and the second spin sign of $u_{2}$ is then fixed by the
condition that the total number of $(-)$ spin sign is odd.

In sum, among $2^{4}\times2^{4}=256$ PFSSA, only $16$ of them are of leading
order in energy in the hard scattering limit. More importantly, all the $16$
leading order amplitudes share the same functional forms and are independent
of the choices of spin polarizations. This result justifies and extends Gross
conjecture \cite{GM,Gross,GrossManes} on high energy string scattering
amplitudes to the fermionic sector.

\section{Discussion}

In contrast to the PFSSA considered in this paper, in the more familiar
palarized fermion field scattering amplitude (PFFSA) calculation in quantum
field theory, the leading order non-vanishing hard (ie. massless limit) PFFSA
are in general NOT proportional to each other. We give two examples here. In
QED, for the lowest order process of $e^{-}e^{+}\longrightarrow\mu^{-}\mu^{+}%
$, there are $4$ non-vanishing among $16$ hard PFFSA \cite{Peskin}%
\begin{align}
\mathcal{M}(e_{R}^{-}e_{L}^{+}  &  \longrightarrow\mu_{R}^{-}\mu_{L}%
^{+})=\mathcal{M}(e_{L}^{-}e_{R}^{+}\longrightarrow\mu_{L}^{-}\mu_{R}^{+}%
)\sim\text{ }(1+\cos\theta)=2\text{ }\cos^{2}\frac{\theta}{2},\\
\mathcal{M}(e_{R}^{-}e_{L}^{+}  &  \longrightarrow\mu_{L}^{-}\mu_{R}%
^{+})=\mathcal{M}M(e_{L}^{-}e_{R}^{+}\longrightarrow\mu_{R}^{-}\mu_{L}%
^{+})\sim\text{ }(1-\cos\theta)=2\text{ }\sin^{2}\frac{\theta}{2},
\end{align}
and they are not all proportional to each other. Note that the usual
\textit{unpolarized} cross section obtained by summing over final spins and
averaging over the initial spins in the hard scattering limit is%
\begin{equation}
\frac{1}{4}\underset{spins}{\sum}\left\vert \mathcal{M}\right\vert ^{2}%
\sim(1+\cos^{2}\theta)\text{ }.
\end{equation}

The second example is the lowest order process of the elastic scattering of a
spin-one-half particle by a spin-zero particle such as $e^{-}\pi
^{+}\longrightarrow e^{-}\pi^{+}$ \cite{JW}. The non-vanishing amplitudes were
shown to be \cite{JW}%
\begin{align}
\mathcal{M}(e_{R}^{-}\pi^{+}  &  \longrightarrow e_{R}^{-}\pi^{+}%
)=\mathcal{M}(e_{L}^{-}\pi^{+}\longrightarrow e_{L}^{-}\pi^{+})\sim\text{
}\cos\frac{\theta}{2},\\
\mathcal{M}(e_{R}^{-}\pi^{+}  &  \longrightarrow e_{L}^{-}\pi^{+}%
)=\mathcal{M}(e_{L}^{-}\pi^{+}\longrightarrow e_{R}^{-}\pi^{+})\sim\text{
}\sin\frac{\theta}{2}.
\end{align}
They are again not all proportional to each other.

This paper is the first attack by the present authors to probe high energy,
higher spin fermion string scatterings. There are many interesting related
issues which remained to be studied. To name a few examples, are there linear
relations among hard fermion SSA so that all the fermion SSA can be solved and
expressed in terms of one amplitude? can these relations be extended to
connect hard SSA of string states of NS sector and R sector ? We will come
back to these interesting topics in the near future.%

%TCIMACRO{\TeXButton{equation number}{\setcounter{equation}{0}
%\renewcommand{\theequation}{\arabic{section}.\arabic{equation}}}}%
%BeginExpansion
\setcounter{equation}{0}
\renewcommand{\theequation}{\arabic{section}.\arabic{equation}}%
%EndExpansion

\begin{acknowledgments}
We would like to thank Y.Okawa for his comments in the early stage of this
work and C.W. Kao for discussion on section IV. This work is supported in part
by the Ministry of Science and Technology (MoST) and S.T. Yau center of
National Chiao Tung University (NCTU), Taiwan.
\end{acknowledgments}

\end{document}